%% file: main.tex
\documentclass[ prl, twocolumn, nofootinbib, floatfix,superscriptaddress]{revtex4-2}

\input{preamble}

\begin{abstract}
    An open question in quantum gravity is how to describe gravitational configurations in terms of quantum states in a Hilbert space. We postulate that quasiclassical states, which are Gaussian, ``coherent state'' distributions over geometries are essential in such descriptions. Working in canonical quantum gravity, we construct a Hamiltonian formalism for gravitational dynamics based on the quasiclassical assumption. As a general feature, one obtains superpositions of non-orthogonal geometry states in which the relative amplitudes redistribute over time. If one constrains the model using semiclassical inputs, the appropriate semiclassical dynamics are recovered, but with additional characteristic quantum features. We apply our formalism to a toy model of black hole evaporation. We also briefly discuss how, in this setting, the high-level description generated by our framework may be constructed from microscopic degrees of freedom.
\end{abstract}

\begin{document}

\input{title_and_authors}

\maketitle
\flushbottom

\textit{Introduction}---The quantization of general relativity has been one of the central objectives of theoretical physics for the past century. Existing frameworks for this quantization can be divided into background-dependent approaches---such as string theory \cite{Polchinski_1998} and perturbative quantum gravity \cite{kiefer2025quantum}---and background-independent approaches---such as canonical quantum gravity \cite{DeWitt1967,dewittPhysRev.162.1195,dewittPhysRev.162.1239,Wheeler1968} and subsequently loop quantum gravity (LQG) \cite{Rovelli_Vidotto_2014}.

Background-independent approaches usually start with the Arnowitt-Deser-Misner (ADM) formulation of general relativity \cite{arnowittPhysRev.116.1322}, whereupon canonical variables are identified and quantized. This process allows one to construct a Hilbert space for observables and states of the theory, although such a construction, including the definition of an inner product for geometries, remains a topic of ongoing investigation. The prototypical example of the canonical program is the Wheeler-DeWitt (WDW) equation, which is given by the Hamiltonian constraint \cite{kiefer2025quantum}
\begin{align} \label{eq:WDW}
    \hat{H}|\Psi\rangle=0,
\end{align}
where the canonical variables are the spatial 3-metric $\hat h_{ab}$ and its conjugate momentum $\hat \pi^{ab}$ (associated with the extrinsic curvature) satisfying \smash{$[\hat h_{ab}(\mathbf x), \hat \pi^{cd}(\mathbf x)] = i \hbar \delta^{c}_{(a} \delta_{b)}^d \delta(\mathbf x , \mathbf y)$}, where $\delta^c_{(a}\delta^d_{b)}$ is the symmetrized Kronecker delta.\footnote{Note that $(\hat h_{ab},\hat\pi^{ab})$ is not the only possible choice of canonical variables, and there are several different schemes for canonically quantizing gravity \cite{kiefer2025quantum}. Here, we use the language of the quantum geometrodynamics approach for the sake of simplicity and clarity, and we stress that our framework is independent of the specifics of the chosen canonical quantization scheme.} The WDW equation tells us that the Hamiltonian annihilates the physical state $| \Psi \rangle$ (a wavefunctional of the 3-metric and momentum), with the implication that $|\Psi\rangle$ does not evolve with respect to a privileged time variable. The absence of dynamical time evolution is referred to as the problem of time \cite{IshamProblemOfTime,unruhPhysRevD.40.2598,kuchardoi:10.1142/S0218271811019347,andersonhttps://doi.org/10.1002/andp.201200147}. 

Due to the technical challenges that arise in the quantization of general relativity \cite{kiefer2025quantum}, we are motivated to ask whether foundational insights into the quantum properties of spacetime can be attained without relying on a full-fledged theory. Recent contributions in quantum foundations and quantum information have shed light on the application of the superposition principle to relativistic spacetime from the operationally motivated perspectives of quantum field theory (QFT) \cite{belenchiaPhysRevD.98.126009,Arrasmith2019,CHRISTODOULOU201964,FooDeSitterThermalMinkowski,FooBTZResonance,FooMinkowski,SuryaatmadjaBTZRotating,Liu2025, Foo2025superpositionsof,foodoi:10.1142/S0218271822420160,GoelAccelerated,akilPhysRevD.108.044051,An2024,kakuPhysRevD.111.046026,Pranzini_2025,pitelli3nfv-lhs1} and quantum reference frames \cite{Castro-Ruiz2020,GiacominiEquivalencePrinciple,DeLaHametteSymmetry,delaHamette2023,Kabel2025,SMoller2024gravitational,foo2025relativitydecoherencespacetimesuperpositions,hohnPhysRevD.109.105011,DEsposito2025doublyquantum}. In these approaches, one takes the mapping $g_{\mu\nu}\mapsto\ket{g_{\mu\nu}}$, without assuming full knowledge of the Hilbert space of geometries. If there are multiple configurations $\{\ket{g_n}\}$ (suppressing coordinate indices),  
then it is commonly assumed that these states are orthonormal.

We posit that the relevant states to consider in quantum gravity are not
the states $\{\ket{g_n}\}$ but rather \emph{quasiclassical} states $\ket{\bar g_n}$. Roughly speaking, quasiclassical states are Gaussian distributions of geometries,   analogous to coherent states found in other areas of quantum physics. We use the notation $\ket{\bar g_n}$ to denote a quasiclassical state centered about $\ket{g_n}$, which in turn corresponds to the classical metric $g_n$ (presently, we overlook the uncertainty of the quasiclassical state, except to remark that it is non-vanishing). From this premise,  previous approaches to spacetime superposition ~\cite{FooDeSitterThermalMinkowski,FooBTZResonance,FooMinkowski,foo2025relativitydecoherencespacetimesuperpositions} can be generalized by recognizing that quasiclassical states are non-orthogonal.

A conceptual challenge in justifying the quasiclassical description is the lack of suitable uncertainty relations pertaining to $g_{\mu\nu}$ arising from well-defined, non-commuting conjugate observables. Canonical quantum gravity and the WDW equation, Eq.~\eqref{eq:WDW}, provide an opportunity to concretize the formalism, since the commutation relations between $\hat h_{ab}$ and $\hat \pi^{ab}$ are known. Henceforth we therefore consider states $\ket{\bar h_n}$  instead of $\ket{\bar g_n}$, where $\ket{\bar h_n}$ is a quasiclassical state about a 3-dimensional spatial metric. However, we will see later that this allows us to describe the dynamics of systems in the abstract Hilbert space of the $\ket{\bar h_n}$'s, after supplying some assumptions.

In this paper, we present a novel framework for computing quasiclassical dynamics. The inputs to this model are a Hamiltonian for the spatial geometry, $\hat H_G$, and the inner products $\langle\bar h_n|\bar h_m\rangle$ for all $m,n$. As a general feature, we observe  redistribution   of quantum amplitude over the $\ket{\bar h_n}$'s: a system that begins in the state $\ket{\bar h_j}$ for some $j$ will, in general, evolve into other quasiclassical geometries. By this, we mean that, although quasiclassical geometries have non-vanishing inner product by construction, an initially exponentially suppressed overlap $\langle \bar h_n|\psi_G(t=0)\rangle$, with $\ket{\psi_G(t)}$ being the state of the spatial system at time $t$, may become dominant in the amplitude at later times, and vice versa.

Another important characteristic of our framework is its ability to fit to phenomenologically motivated inputs while simultaneously predicting novel quantum features. As an example, we consider the evaporation of a black hole, where $\{\ket{\bar h_n}\}$ correspond to spatial descriptions of the black hole at different masses. We show that, if one chooses an appropriate energy spectrum for $\hat H_G$ (namely, a spectrum that corresponds to the Stefan-Boltzmann law), along with suitable choices of $\langle\bar h_n|\bar h_m\rangle$, then our model recovers the semiclassical evaporation relation $M_\mathrm{BH}(t) \sim - t^{1/3} + \mathrm{const.}$ as the trajectory of highest probability within the full quantum wavefunction evolution. However, we find additional quantum interference that goes beyond the semiclassical approximation; this ``extra quantumness'' is a general feature.

In existing approaches to quantum gravity, it is generally difficult to describe dynamical processes. We expect our framework to furnish coarse-grained yardsticks against which other theoretical descriptions may be compared and contrasted. In relying on simple assumptions, we produce a description of spacetime dynamics that connects low-energy quantum information approaches to gravitational quantum physics and top-down canonical frameworks for quantum gravity, while providing a new path for exploring physics beyond QFT in curved spacetimes.

\textit{The Quasiclassical Framework}---Beginning with the WDW equation, which does not have dynamical time evolution, we introduce the decomposition $\hat H=\hat H_G+\hat H_C$, so that Eq. \eqref{eq:WDW} becomes
\begin{align} 
(\hat{H}_G+\hat{H}_C)|\Psi\rangle=0.
\label{eq2}
\end{align}
$\ket{\Psi}$ is the state of a system that is described by the Hilbert space $\mathcal{H}_G \otimes \mathcal{H}_C$, representing the geometric and clock degrees of freedom (DoFs) respectively. $\hat H_G$ acts on the geometric subspace while $\hat H_C$ acts on a clock subspace.

We introduce our clock in the spirit of Page and Wootters ~\cite{PageWootters1,PageWootters2,PageWootters3}. Specifically (and following Ref.\ \cite{SmithClocks}), we associate with the clock Hilbert space $\mathcal{H}_C \simeq L^2(\mathbb R)$ a time operator $\hat T$ that is canonically conjugate to the clock Hamiltonian, $[ \hat T, \hat H_C ] = i$ (we adopt $\hbar=c=G_N=1$). The states of the clock $| t \rangle$ correspond to an instant of time $t$, and we take the clock (for now) to be ideal such that $\langle t | t' \rangle = \delta(t-t')$, implying the spectral decomposition $\hat T = \int\mathrm d t \: t | t \rangle\langle t |$. The physical states $|\Psi\rangle=\int \dd t~|t\rangle|\psi_G(t)\rangle$ are interpreted as an entangled state of the clock with the geometry. One obtains the evolution of the system by conditioning the clock at different times. 
After imposing this constraint in Eq.\ (\ref{eq2}), one finds that the gravitational DoFs evolve via the Schr\"odinger equation,
\begin{align} \label{eq:schroedinger}
    i \frac{\partial}{\partial t} | \psi_G(t) \rangle = \hat H_G | \psi_G(t) \rangle . 
\end{align}

By splitting the total Hamiltonian as in Eq.\ (\ref{eq2}), we assume that the clock system is decoupled from the geometric DoFs. The absence of an interaction term between clock and geometry is a strong restriction, although certain systems---for example, homogeneous cosmologies (either vacuum or with a massless scalar field)---satisfy this decoupled condition \cite{Hoehn:2019fsy}. We may also consider a clock that is guaranteed to be in a region where $\hat H_G\approx 0$---for example, a clock that is situated in an asymptotically flat region that is shared by all possible spatial configurations of the system. More generally, one may consider a clock that couples explicitly to the geometry, leading to modified Schr\"odinger dynamics of the geometry DoFs (see Ref.\ \cite{SmithClocks}, in which the form of the modified Schr\"odinger equation is derived).

We consider, initially, an ideal clock with perfect temporal resolution. Consequently, the clock has unbounded energy, leading to infinite backreaction on the spacetime, although the incorporation of finite time resolution is straightforward and has been developed in e.g.\ Ref.\ \cite{Smith2020,Hoehn:2019fsy}. For simplicity, we assume that $\langle t|t'\rangle \simeq \delta(t-t')$, so that the clock can be approximated as ideal. If we impose that $\hat H_G\approx 0$ in the neighbourhood of the clock, then any finite energy, even if large, would  
yield a finite controllable backreaction.

From Eq. \eqref{eq:schroedinger}, we deduce that only $\hat H_G$ and the inner products $\langle\bar h_n|\bar h_m\rangle$ must be supplied to compute the dynamics for any initial state. Although there are many possible definitions of $\hat H_G$, we present below a natural starting choice. Consider first the case of an orthonormal basis of geometry states, $\{| h_n \rangle\}$, in $\mathcal{H}_G \simeq \mathbb C^{N+1}$. We do not claim to provide a full construction of the Hilbert space for the metric or address the mathematical challenges surrounding the existence of a well-defined measure over the set of all geometries \cite{kiefer2025quantum}---nevertheless, we assume that, in a consistent theory of quantum gravity, such a construction exists. Quantizing the energy $E$ yields
\begin{align}
    \hat H_{G} &= \sum_{n=0}^N E_n | h_n \rangle\langle h_n | , \:\:\: \langle h_n | h_m \rangle = \delta_{nm},
    \label{eq3hamiltonian}
\end{align}
where each $n$ is associated with a macroscopically distinct spatial geometry.

In going from $\ket{h_n}$ to $\ket{\bar h_n}$, we may think of $\ket{\bar h_n}$ as a coherent state peaked about $h_n$. Treating quasiclassical geometries as coherent states follows directly from the quantization of general relativity, the first step being the imposition of the commutator between the canonical variables $\hat h_{ab}$ and $\hat \pi^{ab}$. It follows from the uncertainty principle that no state in the gravitational phase space can be arbitrarily localized. Hence, the relevant states describing classical geometries are coherent states, which minimize the uncertainty with equal variance in each quadrature over this phase space. Indeed, metric eigenstates that satisfy $\hat{h}_{ab}|h\rangle=h_{ab}|h\rangle$ possess maximal uncertainty in $\hat{\pi}^{ab}$ and thus disperse rapidly into a superposition of eigenstates \cite{PapadodimasRaju2016}. This differs from classical general relativity, in which a spacetime is uniquely characterized by $c$-numbers $(h_{ab},\pi^{ab})$.

\begin{figure*}[t]
    \centering
    \includegraphics[width=0.8\linewidth]{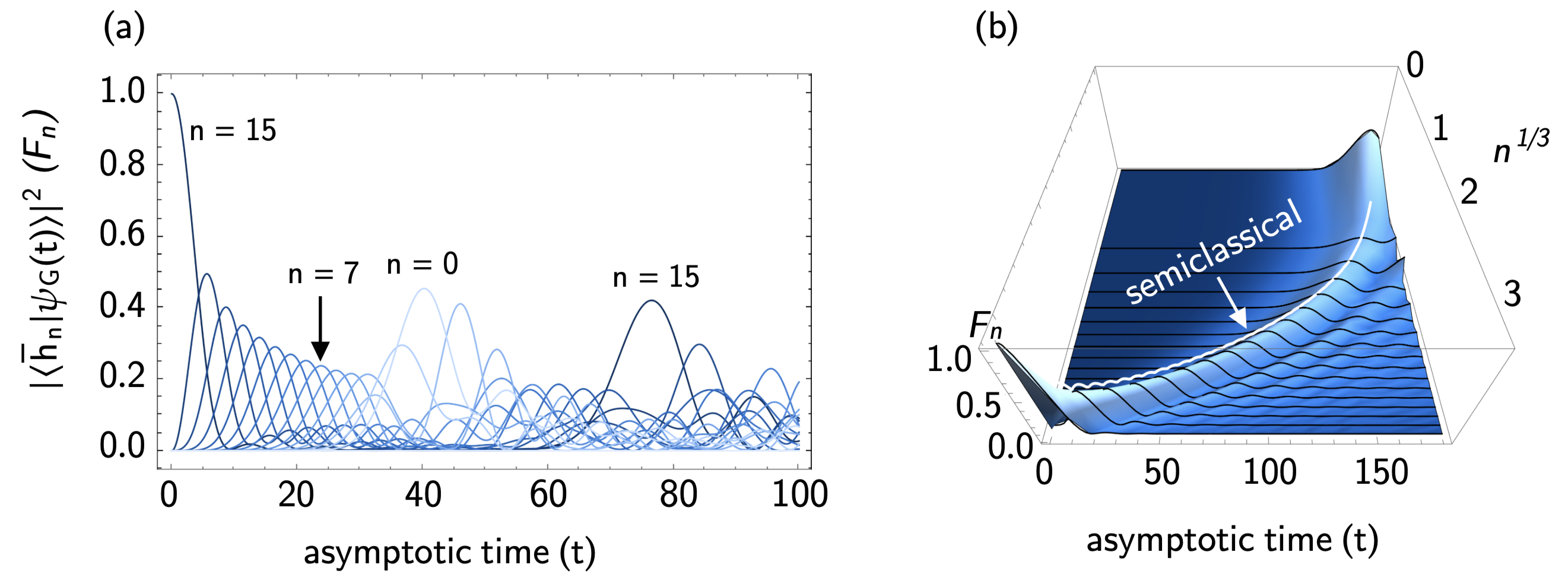}
    \caption{Toy model of black hole evaporation, with the spectrum $\bar E_n = \bar E_0 + \alpha n^{1/3}$. (a) Time-evolution of the fidelity $F_n(t) = | \langle \bar h_n | \psi_G(t) \rangle |^2$ for initial state $| \bar h_{15} \rangle$ and different values of $n$ for the projected state. Here, $N = 15$. (c) $F_n(t)$ plotted as a function of $(n,t)$, where individual black curves correspond to a single value of $n$, with the Hilbert space truncated at $N = 32$, choosing initial state $\ket{\bar h_{32}}$. The white curve is the semiclassical prediction $\bar E_n(t)  = ( \bar E^3_0 - \eta t)^{1/3}$ with $\eta$ a proportionality constant. The semiclassical curve matches the most probable trajectory of the full unitary dynamics (i.e.\ the ridge followed by the blue-white surface). In all plots $\bar E_0 /\alpha = 1/1000$ and $\varepsilon = 1/10$.
    }
    \label{fig:results}
\end{figure*}

With this in mind, we propose the following modification to Eq.\ (\ref{eq3hamiltonian}), 
\begin{align}
    \label{eq4hamiltonian}
    \hat H_G &= \sum_{n=0}^N \bar E_n | \bar h_n \rangle\langle \bar h_n | , \quad \langle \bar h_n | \bar h_m \rangle = e^{-\beta v(n,m)},
\end{align}
which is now expanded in the non-orthogonal basis of gravitational coherent states. We further assume that the inner product is a function of the phase space distance $v(n,m)$ between geometries $| \bar h_n \rangle,~| \bar h_m \rangle$ while $\beta$ is a dimensionless constant that we expect to depend on $\hbar,~c ,~\mathrm{and} ~G_N$. We note that in Ref.\ \cite{PapadodimasRaju2016}, the authors define an inner product similar to Eq.\ (\ref{eq4hamiltonian}) in the linearized gravity regime, with $v(n,m)$ obtained by treating $\bar g_n$ as an excitation over $\bar g_m$ (in the context of that reference, the authors considered spacetime 4-metrics rather than spatial 3-metrics). We consider the form of $\langle \bar h_n | \bar h_m \rangle$ in Eq. \eqref{eq4hamiltonian} to be sufficiently generic for a first choice.

At this point, we take a step back to make several remarks. First, we emphasize that Eq. \eqref{eq4hamiltonian} is a concrete choice of $\hat H_G$, but this choice is by no means integral to the framework that we have proposed. We write Eq. \eqref{eq4hamiltonian} to show that, by making simple generalizations and assumptions, one may define a relatively straightforward Hamiltonian for the quantum geometry, and we use this form of $\hat H_G$ in the later example. Second, we have assumed a discrete set of quasiclassical states for simplicity, but a continuum of states is also possible. Third, if one takes $\beta=-\ln(\ve)$, where $\ve\ll 1$, then $\langle \bar h_n | \bar h_m \rangle = \ve^{v(n,m)}$, which, for a suitable $v(n,m)$, naturally admits a perturbative strategy.

\textit{Reproducing the Semiclassical Evaporation Curve of a Black Hole, An Example}---Let us now consider applying the Hamiltonian, Eq.\ \eqref{eq4hamiltonian}, to analyze the quantum dynamics of a quasiclassical Schwarzschild black hole. The states $\ket{\bar h_n}$ are coarse-grained descriptions of the black hole, each state centered about a different Misner-Sharp mass and representing a possible stage of evolution. Semiclassically, we know that such a black hole will evaporate. We now use our framework to model an analogous  process of evaporation for the Schwarzschild black hole using fully quantum wavefunction evolution. In standard quantum gravity approaches, such a dynamic process would be difficult to characterize. Although $\ket{\bar h_n}$ describes a black hole state in the broadest coarse-grained sense, one may construct more fine-grained models based on $\ket{\bar h_n}$. In Appendix A, we present a black hole toy model involving particles, which satisfies the inner product structure and energy spectrum that we define in this example.

Taking $\beta=-\ln(\ve)$ for $\ve\ll 1$, we define $v(n,m)=|n-m|$, and hence $\langle\bar h_n|\bar h_m \rangle=\ve^{|n-m|}$. This is effectively a nearest-neighbour structure, with more distant overlaps being suppressed by powers of $\ve$. The choice of overlap structure guides, in part, the evolution of the system, although there is some robustness concerning this choice. For example, $v(n,m)=\mathrm{Ceiling}(|n-m|/2)$ (i.e., nearest- and next-to-nearest neighbours) yields qualitatively similar results as $v(n,m)=|n-m|$. More exotic structures would yield different evolutions, possibly unphysical ones, depending on the context in which the model is applied. We believe that the flexibility of our framework, resulting from the generality of its assumptions, is more a strength than a defect; in studying the framework in more contexts, parameter choices can be made with greater precision.

It remains only to define the $\bar E_n$'s in $\hat H_G$, given in Eq. \eqref{eq4hamiltonian}. As previously stated, we take $\bar E_n$ to correspond to the (average) Misner-Sharp mass of the black hole state indexed by $n$. 
This condition alone does not narrow down the possibilities. However, we are free to input empirical data to constrain our model in a realistic manner. Specifically, the energy gap $\Delta\bar{E}_{n}=\bar{E}_n-\bar{E}_{n-1}$ must be related to the thermodynamical time-dependence of the black hole evaporation. Assuming the black hole is a perfect blackbody, then, for consistency with quantum field theory in curved spacetime (QFTCS), the Stefan-Boltzmann law in four spacetime dimensions predicts that $\mathrm dE /\mathrm dt \sim AT^4 \sim 1/ E^2$, where $A$ is the horizon area and $T$ is the black hole temperature. Based on this, we apply our approach to the four-dimensional Schwarzschild black hole. In particular, the spectrum $\bar E_n = \bar E_0 + \alpha n^{1/3}$ ($\alpha$ is a proportionality constant and $\bar E_0$ is a constant offset) gives rise to the finite-difference relation, 
\begin{align}
    \frac{\Delta \bar E_n}{\Delta n} \sim n^{-2/3} \sim \bar E_n^{-2}.
\end{align}
Due to the nearest-neighbour structure of $\langle\bar h_n|\bar h_m \rangle$, $\Delta n$ is a proxy for time (i.e., our system evolves with $\Delta n$ proportional to $\Delta t$). Hence, an $n^{1/3}$ dependence yields the energy spectrum commensurate with the Stefan-Boltzmann law.

With these ingredients, we compute the full dynamics of our system. In Fig. \ref{fig:results}, we perform exact numerical diagonalization of the Hamiltonian, plotting $F_n(t) = | \langle \bar h_n| \psi_G(t) \rangle |^2$ as a function of time, for each $n$. We choose the initial state to be the highest energy state in the truncated Hilbert space, $| \psi_G(0) \rangle = | \bar h_N \rangle$. Since the $| \bar h_n \rangle$'s are approximately orthogonal, the fidelity $F_n(t)$ may be interpreted as a probability distribution up to $O(\ve^2)$ corrections. In Fig. \ref{fig:results}(a), we show the probability amplitude of each $n$ over time as a separate curve, choosing $N=15$ so as not to clutter the plot. Starting from the maximal state $\ket{\bar h_{15}}$, the probability amplitude redistributes to states with lower mean energy. This proceeds until $n=0$ (the lowest energy state), which exhibits a pronounced peak.

In Fig.\ \ref{fig:results}(b), we plot the fidelity $F_n(t)$ for each $n$ in the larger model $N=32$, displaying the results on the same surface. The white curve shows that the trajectory of maximal fidelity reproduces the semiclassical evaporation curve $M_\mathrm{BH}(t) = (M_{0,\mathrm{BH}}^3 - \eta t)^{1/3}$, where $\eta$ is a proportionality constant and $M_{0,\mathrm{BH}}$ is the initial mass. Instead of a sharp ridge, however, we observe fluctuations around and to the right of the ridge of maximal probability. Although our model is fit to a semiclassical expectation, additional quantumness arises from the quasiclassical assumption. This extra quantumness is intrinsic to the quasiclassical framework and is a new feature of quantum spacetime evolution that our approach predicts.


\textit{Conclusion}---Positing that quantum spatial geometry is quasiclassical (i.e., a coherent state), we have proposed a Hamiltonian formalism for spacetime dynamics in the framework of canonical quantum gravity. Our model takes as inputs a Hamiltonian $\hat H_G$ that acts on the geometry and a set of inner product relations $\langle\bar h_n|\bar h_m \rangle$ between quasiclassical states. We fit this model to the semiclassical expectation for an evaporating Schwarzschild black hole, producing a coarse-grained description of the dynamics that can be computed using the Schr\"odinger evolution. This evolution recovers the semiclassical trajectory, but with additional quantum fluctuations. In general, one obtains from our model a redistribution of quantum amplitude over quasiclassical states, along with wave-like dispersion and interference patterns.

We remark that the Hamiltonian in Eq.\ (\ref{eq4hamiltonian}) was not obtained from an underlying microscopic model for the gravitational DoFs, which would require fully-fledged quantum gravity. The strength of the formalism developed here is the ability to analyze generalizations of the form of Eq.\ (\ref{eq4hamiltonian}) without resorting to a full theory. We also note that, while Eq.\ (\ref{eq4hamiltonian}) has several desirable properties---for example, it provides a description of ``fuzzy'' geometries---it is not a unique choice of Hamiltonian or overlaps; the investigation of other possibilities is left for future work.

That being said, our framework is not incompatible with microscopic descriptions. It may even inform such constructions, as we demonstrate in Appendix A. Recent work by Akil et.\ al.\ \cite{akil2025quantumsuperpositionblackhole} adopts a similar mindset to ours; in particular, Ref.\ \cite{akil2025quantumsuperpositionblackhole} describes the coherent evolution of a black hole and Hawking radiation via repeated actions of quantum-controlled unitaries. We will investigate more fine-grained models for black hole evaporation and their implications for the black hole information paradox in a forthcoming paper.

We have also assumed a clock that is, for all practical purposes, ideal and that does not interact with the geometry. An immediate generalization of our model is to relax these assumptions. We also, for simplicity, consider a discrete spectrum for $\hat H_G$, which leaves the continuous case as a clear next step. The problem of measurement for spacetime superposition must also be addressed. We remark that the non-orthogonality of the quasiclassical states is commensurate with the idea that a classical parameter, such as the mass of a black hole, cannot be estimated with arbitrary precision.

The late-time dynamics of our model also warrant further investigation. After the peak corresponding to $\ket{\bar h_0}$ is obtained in Fig. \ref{fig:results}(a), the probability reflects back to higher energy states, with increasingly non-trivial interference. However, since $\ket{\bar h_0}$ is a fully evaporated state, one would have to cross the Planckian threshold in order to access the dynamics after the reflection, which raises the question of how one should interpret these late-time predictions. Overall, we believe that there is considerable room to expand our framework, which links low-energy and bottom-up approaches to quantum gravity to top-down strategies, particularly, canonical quantum gravity.

\textit{Acknowledgements}---We thank Maximillian Lock, Nicola Pranzini, Carlo Rovelli, Alexander R.H. Smith, Jiro Soda, and Francesca Vidotto for helpful discussions during the development of this work, and Timothy Ralph for providing critical feedback on an earlier version of this manuscript. SW, RBM, and JF acknowledge support by the Natural Sciences and Engineering Research Council of Canada. AS acknowledges support by the Australian Research Council Centre of Excellence for Quantum Computation and Communication Technology (Project No. CE170100012). JF receives support from a Banting Postdoctoral Fellowship. DY was supported by the National Research Foundation of Korea (NRF) grant funded by the Korean government (No.~RS-2026-25476711).

\bibliography{ref}

\appendix

\section*{Appendices}

\section{Appendix A: Toy Model for Black Hole Evaporation}\label{sec:appendix-a}

In the main text, we proposed  Eq. \eqref{eq4hamiltonian}, which reads
\begin{align}
    \hat H_G &= \sum_{n=0}^N \bar E_n | \bar h_n \rangle\langle \bar h_n | , \quad \langle \bar h_n | \bar h_m \rangle = e^{-\beta v(n,m)}, \nonumber
\end{align}
and we assumed $\beta=-\ln(\ve)$ and $v(n,m)=|n-m|$. By choosing $\bar E_n$ to satisfy the Stefan-Boltzmann law, we showed that it is possible to recover the semiclassical evaporation curve for the time evolution of a Schwarzschild black hole, plus addtional quantum features. In general, our framework predicts that spatial geometry redistributes in amplitude over the basis of quasiclassical configurations.

In this Appendix, we consider a toy model where we explicitly account for particles in the evaporation process. That is, we wish to interpret the results of Fig.\ \ref{fig:results} in terms of a microscopic model of the dynamics where $\ket{\bar h_n}$ includes both the geometry plus ingoing Hawking radiation, but not outgoing Hawking radiation. Tracing out the outgoing modes allows the putative black hole to ``evaporate'' (i.e., to reach $\bar h_0$). 

Let us consider the geometry plus Hawking radiation in the Hilbert space $\mathcal{H}_G \otimes \mathcal{H}_R$ (we neglect the clock for clarity here). We define the non-orthogonal basis states of the $i$th mode as
\begin{align}
    | 1 \rangle_i &= \begin{pmatrix}
             1 \, \\  0 \, \\  0 \, 
    \end{pmatrix}, 
     \:\:\: 
    | 0 \rangle_i = \textsf{N}
    \begin{pmatrix}
         \varepsilon \, \\  1 \, \\  \varepsilon \,
    \end{pmatrix}, \:\:
    | - 1 \rangle_i = \begin{pmatrix}
         0 \, \\  0 \, \\  1\,
    \end{pmatrix},
\end{align}
which respectively correspond to particle, vacuum, and antiparticle states, $\textsf{N} = ( 1 + 2 \varepsilon^2 )^{-1/2}$ is a normalization constant, and we define a local Hamiltonian for the $i$th subsystem as $\hat H_i = \omega ( | 1 \rangle\langle 1 | - | - 1 \rangle\langle -1  | )_i$. As such, the relations $\langle n | \hat H_i | n \rangle = n \omega$ and $\langle \pm 1 | 0 \rangle = \langle 0 | \pm 1 \rangle \simeq \varepsilon$ are satisfied, for $n = 0 , \pm 1$ and for each subsystem $i$. Let us now consider the geometry as being constructed from $N$ modes in the tensor product $| S \rangle = \bigotimes_{i=1}^N | 1 \rangle_i$, representing the star interior. Meanwhile, the radiation is comprised of $M$ separable pairs of particles and antiparticles,  
\begin{align}
    | R \rangle &= \left(\bigotimes_{j=1}^M | - 1 \rangle_{N+2j - 1} \otimes | 1 \rangle_{N+2j}\right) \otimes | F \rangle,
\end{align} 
where 
\begin{align} 
    | F \rangle = \bigotimes_{k=1}^{N-M} | 0 \rangle_{N+2M +2k-1} \otimes | 0 \rangle_{N+2M+2k} 
\end{align}
is the subsystem representing vacuum fluctuations that supplies the non-orthogonality condition needed to induce transitions in the full quantum state $| \psi \rangle = | S \rangle \otimes | R \rangle$ for $M < N$. We assume that the initial state corresponds to $M=0$ such that there is a classical star without Hawking pairs. As time progresses, Hawking pairs are created and $M$ increases. Denoting $| \psi \rangle = |\tilde n\rangle $ with $\tilde n \equiv N - M$, it can be easily shown that $\langle \tilde m | \tilde n \rangle \simeq \varepsilon^{2 | \tilde{n} - \tilde{m} |}$.

Let us now construct the Hamiltonian. First defining $\hat{\mathcal{H}}_i = \mathbb I^{\otimes (i-1)} \otimes \hat H_i \otimes \mathbb I^{\otimes (3N-i)}$, then we can express the joint Hamiltonian of all subsystems as 
\begin{align}
    \hat{\mathcal{H}} &= \sum_{i=1}^{3N}  \hat{\mathcal{H}}_i,
\end{align}
such that the total energy is manifestly conserved: $\langle \tilde n | \hat{\mathcal{H}} | \tilde n \rangle = N \omega$. We may now consider a reduced Hamiltonian, which captures the ingoing parts only:
\begin{align}
    \hat{\mathcal{H}}' &= \sum_{\tilde n=1}^N E_{\tilde n} | \tilde n \rangle\langle \tilde n | 
    = \sum_{i=1}^N \hat{\mathcal{H}}_i + \sum_{j=1}^M \hat{\mathcal{H}}_{N+2j-1},
\end{align}
such that $\langle \tilde n | \hat{\mathcal{H}}' | \tilde n \rangle = E_{\tilde n} = \tilde n \omega$. The ingoing part consists of the star interior plus antiparticles (plus half of the total vacuum modes). Hence, $E_{\tilde n}$ should correspond to the quasilocal mass estimating the energy of the interior, for example the Misner-Sharp mass \cite{misnerPhysRev.136.B571}. $\hat{\mathcal{H}}'$ only acts on the subspace of the $2N$ ingoing DoFs; thus, we may trace out the outgoing modes and express $\hat{\mathcal{H}}'\simeq \sum_{\tilde n=1}^N E_{\tilde n} | \tilde n' \rangle\langle \tilde n' | $, where
\begin{align}
    | \tilde n' \rangle &= | S \rangle \bigotimes_{j=1}^M | -1  \rangle_{N+2j-1} \bigotimes_{k=1}^{N-M} | 0 \rangle_{N+M+2k-1}.
\end{align}
$\hat{\mathcal{H}}'$ realizes the condition in Eq.~(\ref{eq4hamiltonian}), with the analogy that $|\tilde n'\rangle \leftrightarrow |\bar h_n\rangle$ and $E_{\tilde n}\leftrightarrow \bar E_n$. Therefore,  $i\partial_t | \psi' \rangle \simeq \hat{\mathcal{H}}' | \psi' \rangle$ whenever $| \psi' \rangle = \sum_{\tilde n'} c_{\tilde n'}(t) | \tilde n' \rangle$, with $c_{\tilde n'}(t)$ being the coefficients of each state. 

We note that the energy gap $\Delta E_{\tilde n}/\Delta \tilde n'=\omega$ is proportional to $\mathrm d E/\mathrm d t$ according to the two-dimensional Stefan-Boltzmann law. Since each particle can have two (particle or antiparticle) states, for a given star interior state with $N$ particles, the entropy is $S = N \log 2$, while the total energy is $E = N\omega$. Therefore, the Hawking temperature $T = \mathrm d E/ \mathrm d S$ is a constant; if we apply this to the two-dimensional Stefan-Boltzmann law, $\mathrm d E/\mathrm d t \sim T^{2} = \mathrm{const.} \sim \Delta E_{\tilde n}/\Delta \tilde n'$. To summarize, in this simple construction, we obtain the Stefan-Boltzmann proportionality, which is the energy spectrum for $\hat H_G$ that we assume in the main text.


One aspect of our model is that the energy of the system with Hilbert space $\mathcal{H}_G$ can change, if the Hamiltonian $\hat H_G$ is not degenerate. For example, in Fig. \ref{fig:results}(b), the system redistributes towards states with lower energy. This means that for non-degenerate $\hat H_G$, there is a natural partition of interior and exterior systems between which there is exchange of energy, with $\ket{\bar h_n}$ being the interior part. As an additional subtlety, the evolution of the interior state $\ket{\psi_G(t)}$ is governed by the Schr\"odinger equation, meaning that it is unitary. Hence, the loss or gain of energy appears as a  puzzling result. A possible implication is that there exists an anti-universe. Under this construction, the interior state of the mirror universe gains or loses antiparticles rather than particles (or vice versa) as in the regular universe. Hence, total energy is conserved. An investigation of the energy problem in microscopic models will be considered in a forthcoming paper. 

\end{document}

%% file: preamble.tex
\pdfoutput=1 
\usepackage[english]{babel}
\usepackage[T1]{fontenc} 
\usepackage{amsmath,amssymb,amsfonts,amsthm}
\usepackage{graphicx}
\usepackage[usenames,dvipsnames]{color}
\usepackage{enumitem}
\usepackage{verbatim}
\usepackage[percent]{overpic}
\usepackage{rotating}
\definecolor{blueprl}{RGB}{46,48,146}
\usepackage[pdftex,colorlinks=true,urlcolor=blueprl,citecolor=blueprl,linkcolor=blueprl]{hyperref}
\usepackage{array}
\usepackage{mathtools}
\usepackage{lmodern}
\usepackage[normalem]{ulem}
\usepackage{soul} 
\usepackage{braket}
\usepackage{tensor}
\usepackage{dsfont}
\usepackage{bm}
\usepackage{tikz}

\usepackage{mathrsfs}

\usetikzlibrary{decorations.pathmorphing}
\usepackage{CJKutf8}

\definecolor{patriarch}{rgb}{0.5, 0.0, 0.5}
\definecolor{darkraspberry}{rgb}{0.53, 0.15, 0.34}
\definecolor{brinkpink}{rgb}{0.98, 0.38, 0.5}
\definecolor{skobeloff}{rgb}{0.0, 0.48, 0.45}
\definecolor{mypink}{RGB}{226,68,130}
\definecolor{darkpastelgreen}{rgb}{0.01, 0.75, 0.24}
\definecolor{pigmentgreen}{rgb}{0.0, 0.65, 0.31}

\usetikzlibrary{positioning}
\usetikzlibrary{calc,through,backgrounds}

\usepackage{bbm}

\newcommand{\dd}{\text{d}}

\newcommand{\ve}{\varepsilon}

%% file: title_and_authors.tex
\title{Quantum coherent dynamics of quasiclassical spacetimes}

\author{Sijia Wang}
\email{s676wang@uwaterloo.ca}
\affiliation{Department of Physics and Astronomy, University of Waterloo, Waterloo, Ontario, N2L 3G1, Canada}
\affiliation{Waterloo Centre for Astrophysics, University of Waterloo, Waterloo, Ontario, N2L 3G1, Canada}

\author{Achintya Sajeendran}
\affiliation{Centre for Quantum Computation and Communication Technology, School of Mathematics
and Physics, The University of Queensland, St. Lucia, Queensland, 4072, Australia}

\author{Dong-han Yeom}
\affiliation{Department of Physics Education, Pusan National University, Busan 46241, Republic of Korea}
\affiliation{Research Center for Dielectric and Advanced Matter Physics, Pusan National University, Busan 46241, Republic of Korea}
\affiliation{Leung Center for Cosmology and Particle Astrophysics, National Taiwan University, Taipei 10617, Taiwan}
\affiliation{Department of Physics and Astronomy, University of Waterloo, Waterloo, Ontario, N2L 3G1, Canada}
\affiliation{Perimeter Institute for Theoretical Physics,  Waterloo, Ontario, N2L 2Y5, Canada}

\author{Robert B. Mann}
\affiliation{Department of Physics and Astronomy, University of Waterloo, Waterloo, Ontario, N2L 3G1, Canada}
\affiliation{Waterloo Centre for Astrophysics, University of Waterloo, Waterloo, Ontario, N2L 3G1, Canada}
\affiliation{Institute for Quantum Computing, University of Waterloo, Waterloo, Ontario, N2L 3G1, Canada}
\affiliation{Perimeter Institute for Theoretical Physics,  Waterloo, Ontario, N2L 2Y5, Canada}

\author{Joshua Foo}
\email{jfoo@uwaterloo.ca}
\affiliation{Department of Physics and Astronomy, University of Waterloo, Waterloo, Ontario, N2L 3G1, Canada}
